\documentclass[usenatbib]{mn2e}
\usepackage{graphicx,natbib,amssymb,aas_macros}

\def\degg{\hbox{$\null^\circ$\hskip-3pt.}}

\title[DDO210, the Aquarius dwarf]{The stellar content of the isolated transition dwarf galaxy DDO210\thanks{Based in part on data collected at
Subaru Telescope, which is operated by the National Astronomical
Observatory of Japan}}

\author[McConnachie et al.] {Alan W. McConnachie$^1$\thanks{E-mail: alan@uvic.ca}, Nobuo Arimoto$^2$, Mike Irwin$^3$ \& Eline Tolstoy$^4$\\
$^1$Department of Physics and Astronomy, University of Victoria, Victoria, B.C., V8P 1A1, Canada\\
$^2$National Astronomical Observatory of Japan, 2-21-1 Osawa, Mitaka, Tokyo 181-8588, Japan\\
$^3$Institute of Astronomy, University of Cambridge, Madingley Road, Cambridge, CB3 0HA, U.K.\\
$^4$Kapteyn Institute, University of Groningen, Postbus 800, 9700AV Groningen, the Netherlands}

\begin{document}

\maketitle

\begin{abstract}
We use Subaru Suprime-Cam and VLT FORS1 photometry of the dwarf galaxy
DDO210 to study the global stellar content and structural properties
of a transition-type galaxy (with properties intermediate between
dwarf irregular and dwarf spheroidal systems). This galaxy is
sufficiently isolated that tidal interactions are not likely to have
affected its evolution in any way. The colour-magnitude diagrams of
DDO210 show a red giant branch (RGB) population (with an RGB bump), a
bright asymptotic giant branch population, a red clump, young main
sequence stars and blue-loop stars. The youngest stars formed within
the last 60\,Myrs and have a distinct radial distribution compared to
the main population. Whereas the overall stellar spatial distribution
and HI spatial distribution are concentric, the young stars are offset
from the center of DDO210 and are coincident with a `dent' in the HI
distribution. The implied recent star formation rate required to form
the young population is significantly higher than the derived current
star formation rate, by a factor of $> 10$.

Most of the stars in DDO210 are found in a red clump, and its mean
$I$-band magnitude suggests that the majority of stars in DDO210 have
an average age of $4^{+2}_{-1}$\,Gyr. Given this age, the colour of
the RGB implies a mean metallicity of [Fe/H] $\simeq -1.3$. By
comparing the shape of the red clump with models for a variety of star
formation histories, we estimate that an old ($\ge 10$\,Gyr) stellar
population can contribute $\sim 20 - 30$\% of the stars in DDO210 at
most. The unusual star formation history of DDO210, its low mass
estimate and its isolated nature, provide insight into how star
formation proceeds in the lowest mass, unperturbed, dwarf galaxy
haloes.
\end{abstract}

\begin{keywords}
galaxies: dwarf ---  galaxies: individual (DDO210/Aquarius) --- Local Group ---  galaxies: stellar content ---  galaxies: structure
\end{keywords}

\section{Introduction}

The stellar, chemical and structural evolution of dwarf galaxies is
believed to depend strongly on environment. In the Local Group, dwarf
spheroidal (dSph) galaxies are preferentially found as satellites to
M31 and the Milky Way (MW), whereas dwarf irregular (dIrr) galaxies
are preferentially found in isolated locations. This was first
highlighted by \cite{einasto1974} (see also
\citealt{vandenbergh1999}), and implies that the presence of a nearby
large galaxy causes morphological changes in its satellites, primarily
via the loss of gas by a combination of tidal stirring and
ram-pressure stripping
(\citealt{mayer2001a,mayer2001b,grebel2003}). Tidal effects from the
large galaxies may also be responsible for star formation episodes in
their satellites, by triggering compression of their gas. However, the
orbital properties of the MW satellites are difficult to determine,
and so correlations between star formation histories (SFHs) and
orbital phase have yet to be robustly demonstrated.

The stellar evolution of dwarf galaxies is also interesting from a
cosmological viewpoint, in particular regarding the missing satellites
problem. A long standing problem faced by theories of galaxy formation
in the cold dark matter paradigm is that purely gravitational
simulations predict that there should be around an order of magnitude
more dwarf galaxies in the Local Group, particularly around the MW and
M31 (\citealt{kauffmann1993,klypin1999,moore1999}). As recent results
are demonstrating, it appears as if there are numerous low surface
brightness dwarf galaxies which have yet to be discovered
(\citealt{zucker2004a,zucker2006a,zucker2006b,willman2005}; see also
\citealt{willman2006,belokurov2006}) although it is unclear if these
discoveries will increase the observed number by the required order of
magnitude. One potential solution to this difficulty is that not all
dwarf galaxy-scale dark matter haloes can accrete and retain gas to
form stars, and instead reionisation can prevent certain haloes from
forming a luminous component
(\citealt{bullock2000,kravtsov2004,ricotti2005,gnedin2006}). \cite{susa2004}
show that star formation is suppressed significantly by the effects of
reionisation, but that sufficiently massive dwarf galaxies are able to
form stars eventually because the baryons are self-shielded.

Even in these cosmological scenarios, the influence of tidal effects
from the MW and M31 can play important roles in determining the
evolution of a dwarf galaxy. Indeed, tidal effects present one of the
greatest uncertainties in our interpretation of dwarf galaxy
evolution, and it is therefore of considerable interest to study dwarf
galaxies unaffected by the presence of an external tidal field. DDO210
($20^{\rm h}\,46^{\rm m}\,51.8^{\rm s}$, $-12^\circ\,50'\,53''$) is
one of the most isolated galaxies in the Local Group. It was
discovered at the limits of detectability on the Palomar Sky Survey by
\cite{vandenbergh1959}, in the constellation of
Aquarius. \cite{fisher1975} detected it in HI 21cm emission, and later
assigned it an arbitrary distance modulus of $\left(m - M\right) =
25$\,mags, based upon some general similarities to the Local Group
galaxy WLM. For a long time DDO210 remained very poorly studied, and
\cite{marconi1990} were the first to resolve its individual stars, to
a limiting magnitude of $V = 23.5$\,mags.

Despite the lack of any distance measurement, DDO210 was considered to
be a member of the Local Group from soon after its discovery
(eg. \citealt{yahil1977,vandenbergh1979,devaucouleurs1983}). \cite{greggio1993}
were the first to estimate an independent distance to this galaxy,
$\left(m - M\right) \simeq 28$\,mags ($d \simeq 4.3$\,Mpc), placing it
well beyond the zero velocity surface of the Local Group. However,
more recent distance measurements by \cite{lee1999} (L99),
\cite{karachentsev2002} (K02) and \cite{mcconnachie2005a} (M05), all
of which are based upon the tip of the red giant branch (TRGB;
\citealt{lee1993a}), show that DDO210 is, in fact, located at $\sim
1$\,Mpc from the Milky Way (MW), on the periphery of the Local
Group. This places DDO210 at slightly over 1\,Mpc from M31. The
free-fall timescale of DDO210 into M31 or the MW is

\begin{equation}
t_{ff} \simeq
14.4$\,Gyr\,$\left(\frac{r}{1\,{\rm Mpc}}\right)^{\frac{3}{2}}\,\left(\frac{M}{10^{12}\,{\rm M_\odot}}\right)^{-\frac{1}{2}}~,
\label{tff}
\end{equation}

\noindent ie. of order a Hubble time. It is clearly not, and never has
been, a satellite of either of the dominant members of the Local
Group.

\cite{mateo1998a} classifies DDO210 as a transition-type galaxy, with
properties intermediate to those which are usually ascribed to dSph
and dIrr galaxies. Its magnitude is $M_V \simeq -10.6$\,mags (L99),
making it the faintest dwarf galaxy in the Local Group with a
significant gaseous component. Despite the presence of this gas, an
$H\alpha$ study of this galaxy (\citealt{vanzee1997}) shows that there
is no current star formation. This is rather unusual, given that L99
show that this galaxy contains a significant number of young stars,
and that the central portion of this galaxy has undergone a recent
enhancement in its star formation rate (SFR) in the last few hundred Myrs.

\cite{begum2004} conducted deep, high velocity resolution HI 21cm
observations of DDO210 with the Giant Meterwave Radio Telescope. They
find that the velocity field of this galaxy is quite regular, and has
a corrected peak rotation velocity of $\sim 16$\,km\,s$^{-1}$. When
incorporated into mass models of DDO210, this implies a total
dynamical mass of order $10^{7 - 8}$\,M$_\odot$, comparable to many of
the dwarf satellites of M31 and the MW.  Since DDO210 is so isolated,
it offers an excellent opportunity to study the stellar evolution of a
dwarf galaxy in the absence of any significant external tidal fields.

In this paper, we analyse deep, global, multi-colour photometry of
DDO210, obtained with Subaru Suprime-Cam and VLT FORS1. We analyse the
stellar content, structural properties and SFH of this galaxy, and
show that it is dominated by an intermediate age stellar population
with an average age of $\sim 4$\,Gyr. In Section~2, we discuss our
Subaru Suprime-Cam and VLT FORS1 photometric datasets and the
data-reduction procedure. In Section 3, we present deep, global, CMDs
of the stellar populations of DDO210, and use these to derive the
properties of its multiple stellar populations. We compare the CMDs
with synthetic models of various SFHs and use this to estimate the
possible contribution from an ancient ($\gtrsim 10$\,Gyr) stellar
component. In Section 4, we analyse the global structure of DDO210, in
terms of its integrated properties and the radial profiles of the
various stellar populations, and we compare these to the global
distribution of HI from previous Very Large Array (VLA) observations
by \cite{young2003}. We discuss the nature of DDO210 in light of these
new results in Section 5, and we summarise our findings in Section
6. We assume an extinction value towards DDO210 of $E\left(B -
V\right) = 0.052$ (\citealt{schlegel1998}).

\section{The photometric datasets}

\subsection{Subaru Suprime-Cam}

\begin{figure*}
  \begin{center}
    \includegraphics[angle=270, width=8.5cm]{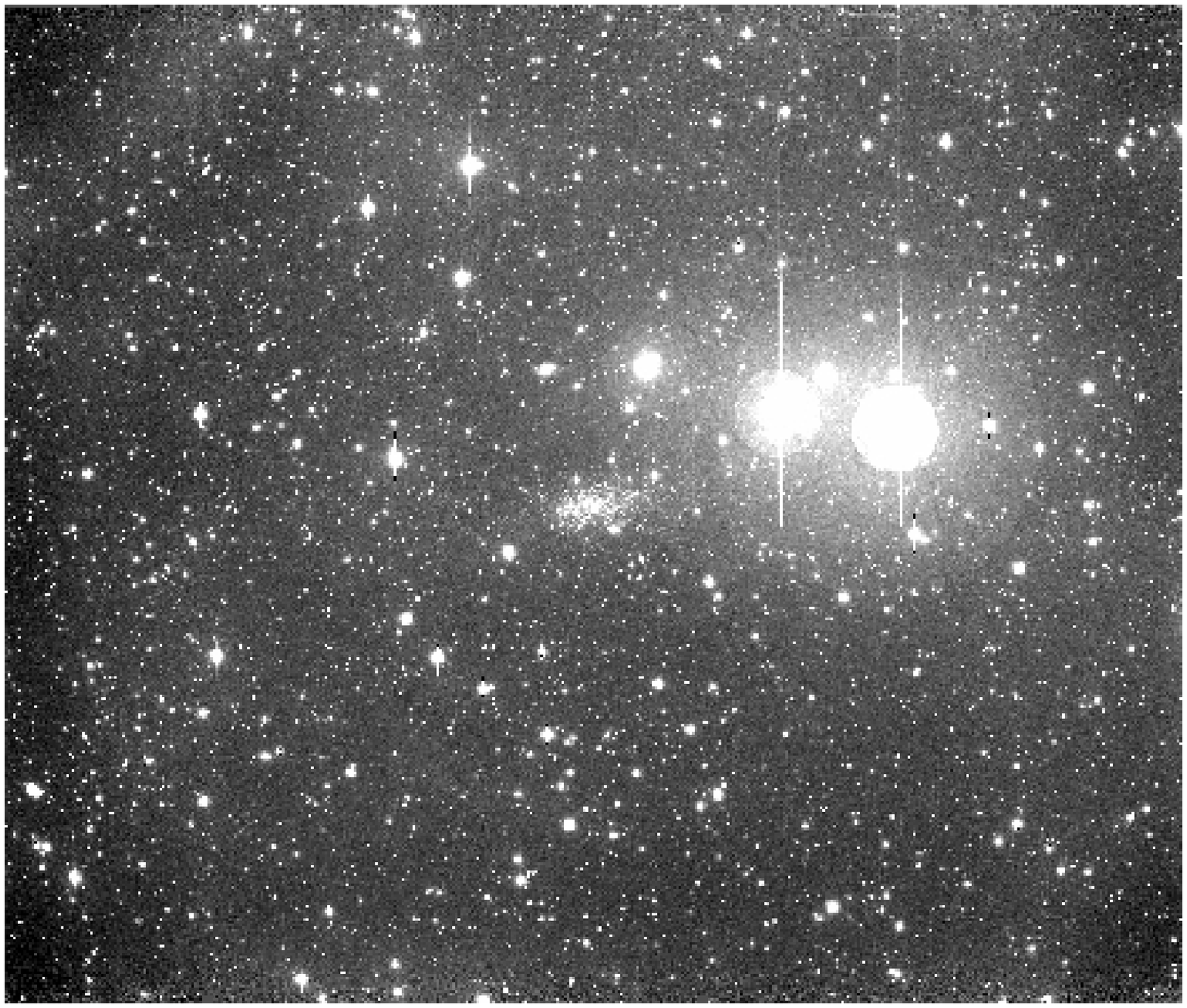}
    \includegraphics[angle=270, width=8.479cm]{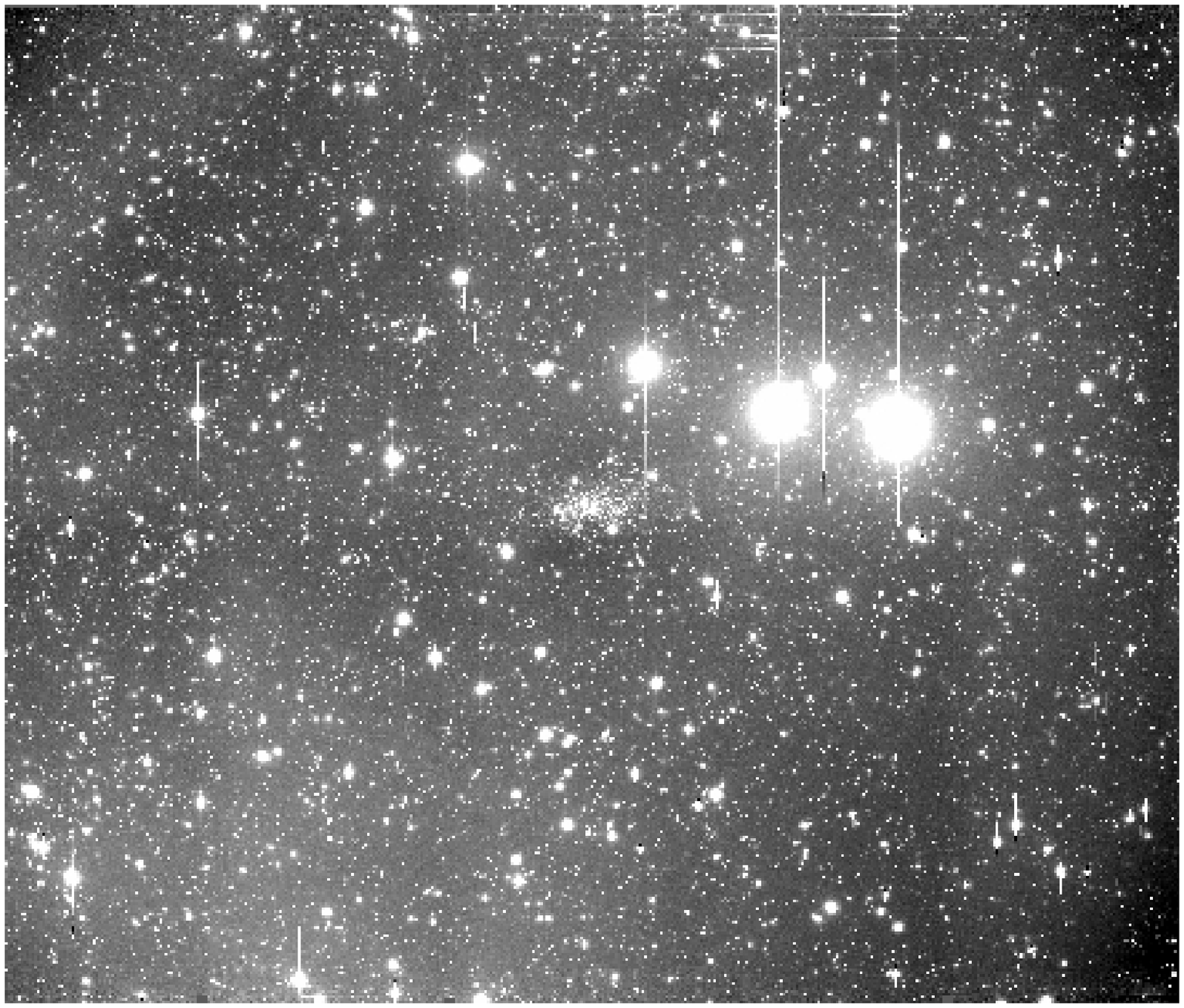}
    \caption{The $V$- and $I_c$-band Suprime-Cam fields,
    centered on DDO210. Each field measures 34\,arcmins $\times
    27$\,arcmins. North is to the top, and east is to the left.}
    \label{image}
  \end{center}
\end{figure*}

During the nights of $3^{{\rm rd}}$ -- $5^{{\rm th}}$ August 2005, we
observed with the $34$\,arcmins $\times 27$\,arcmins Suprime-Cam wide
field camera on the Subaru Telescope obtaining Johnson -- Cousins $V$
and $I_c$-band imaging of several Local Group dwarf galaxies which are
not satellites of the Milky Way (see McConnachie, Arimoto \& Irwin
2006, {\it in preparation}).  Conditions were uniformly excellent,
being photometric throughout and with typical seeing of
$0.5$\,arcsecs. For DDO210, we exposed for a total of $4320$\,seconds
in $V$ and 7200\,seconds in $I_c$, split as $12 \times 360$\,seconds
and $30 \times 240$\,seconds dithered sub-exposures respectively.  The
telescope was typically offset $\sim 20$\,arcsecs between
sub-exposures. Although the $V$-band DDO210 observations were taken at
relatively high airmass ($1.5 - 2.2$), the final stacked images still
have sub-arcsec seeing, averaging $0.82$\,arcsec over the whole array.
For the $I_c$-band images, mostly taken at lower airmass, the seeing
in the stacked image averages $0.51$\,arcsec. The reader is referred
to McConnachie, Arimoto \& Irwin (2006) {\it in preparation} for
details of the data reduction process.

We cross-correlate the Suprime-Cam photometry with our earlier
multi-colour Isaac Newton Telescope Wide Field Camera photometry of
DDO210 (M05), for which we know the colour transformations into the
Landolt
system\footnote{http://www.ast.cam.ac.uk/~wfcsur/technical/photom/colours}. This
ensures our new photometry is on the same systems as our previous
photometry. By only considering those objects reliably identified as
stellar in all four sets of observations, we find

\begin{eqnarray}
V & = & V^\prime + 0.030 \left(V - I\right)\nonumber\\
I & = & I_c - 0.088 \left(V - I\right)~,
\label{sub2lan}
\end{eqnarray}

\noindent where we now use $V^\prime$ to denote the original Subaru
$V$ filter. The transformations are small, and we apply these colour
equations to our Suprime-Cam data.

We have compared our $VI$ photometry to the photometry of L99, for
bright stars with $V < 22$ over their entire field. Based upon tight
spatial coincidence, we confidently identify $26$ common stars, which
have a mean $\Delta\,V = 0.038$ (our photometry minus L99
photometry). This is small in comparison to the various sources of
error; for example, the calibration error in the photometry of L99 is
given as $0.04 - 0.05$\,mags, and the uncertainty in the zero-point of
our photometry is typically $\sim 0.02$\,mags. Thus we conclude that
our photometry is in good agreement with L99, which has itself been
shown to agree well with the earlier studies of \cite{greggio1993} and
\cite{hopp1995}.

\subsection{VLT FORS1}

We have also included in our analysis deep $B$ and $R$-band imaging of
the center of DDO210 taken during excellent stable photometric
conditions in August 1999 with the newly commissioned FORS1 instrument
on UT1 at the VLT (P.I. E. Tolstoy). These filters were chosen to
match the wavelength range in which the instrument is most sensitive
and preliminary results from these observations have been presented in
\cite{tolstoy2000}. The data were originally reduced using standard
procedures, and the stacked $B$ and $R$-band images were subsequently
processed in the same way as the Subaru data. This produced $B$ and
$R$-band detected objects and a combined $B,R$ catalogue.

\section{The stellar populations of DDO210}

\begin{figure}
  \begin{center}
    \includegraphics[angle=0, width=8cm]{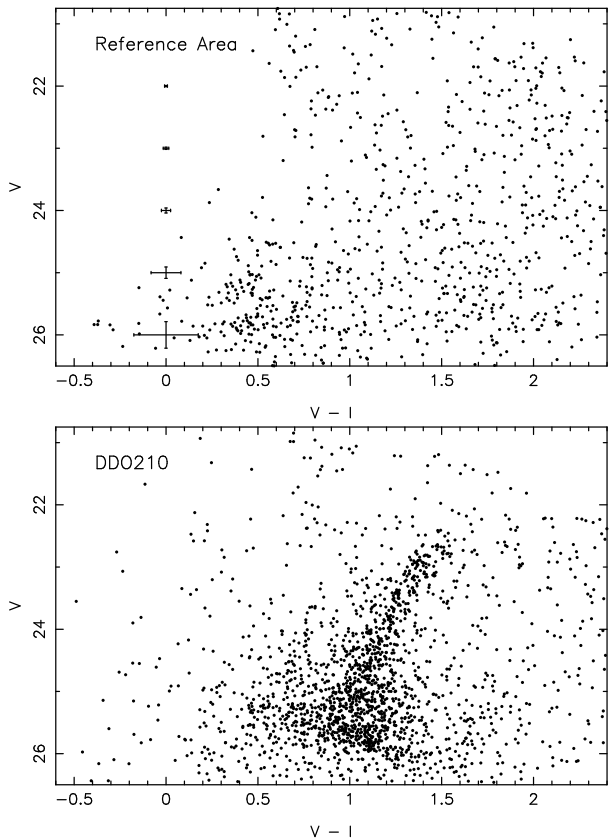}
    \caption{$V$ versus $\left(V - I\right)$ CMD for the
    reference area (top) and for DDO210 (bottom), from our Subaru
    Suprime-Cam imaging. The reference area is located outside of the
    innermost $7.5$\,arcmins $\times 7.5$\,arcmins of the Suprime-Cam
    field, and the DDO210 CMD consists of those stars in the
    innermost $4$\,arcmins $\times 4$\,arcmins.}
    \label{vcmd}
  \end{center}
\end{figure}

\begin{figure}
  \begin{center}
    \includegraphics[angle=0, width=8cm]{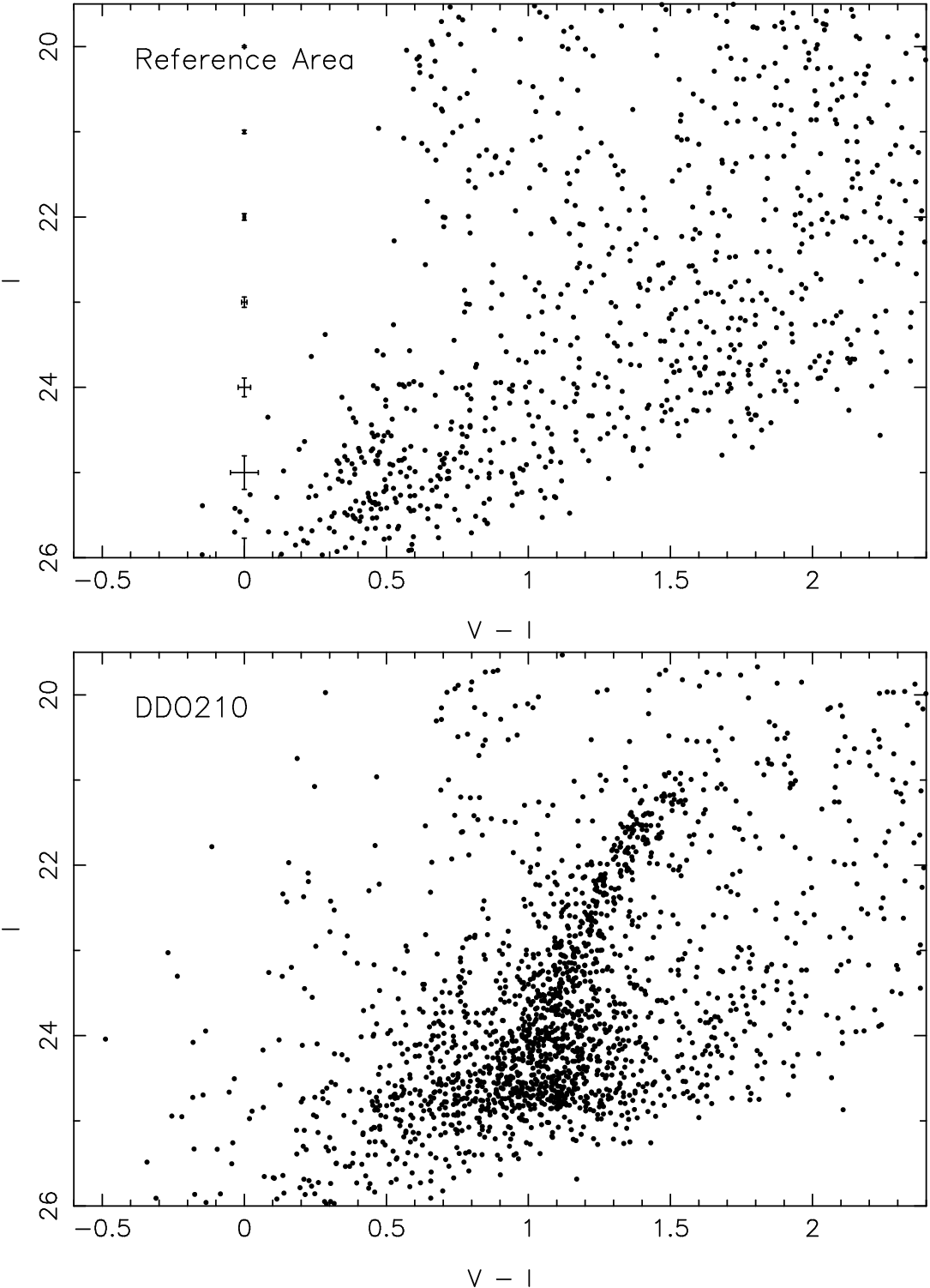}
    \caption{$I$ versus $\left(V - I\right)$ CMD for the
    reference area (top) and for DDO210 (bottom), from our Subaru
    Suprime-Cam imaging. The reference area is located outside of the
    innermost $7.5$\,arcmins $\times 7.5$\,arcmins of the Suprime-Cam
    field, and the DDO210 CMD consists of those stars in the
    innermost $4$\,arcmins $\times 4$\,arcmins.}
    \label{icmd}
  \end{center}
\end{figure}

\begin{figure}
  \begin{center}
    \includegraphics[angle=270, width=8cm]{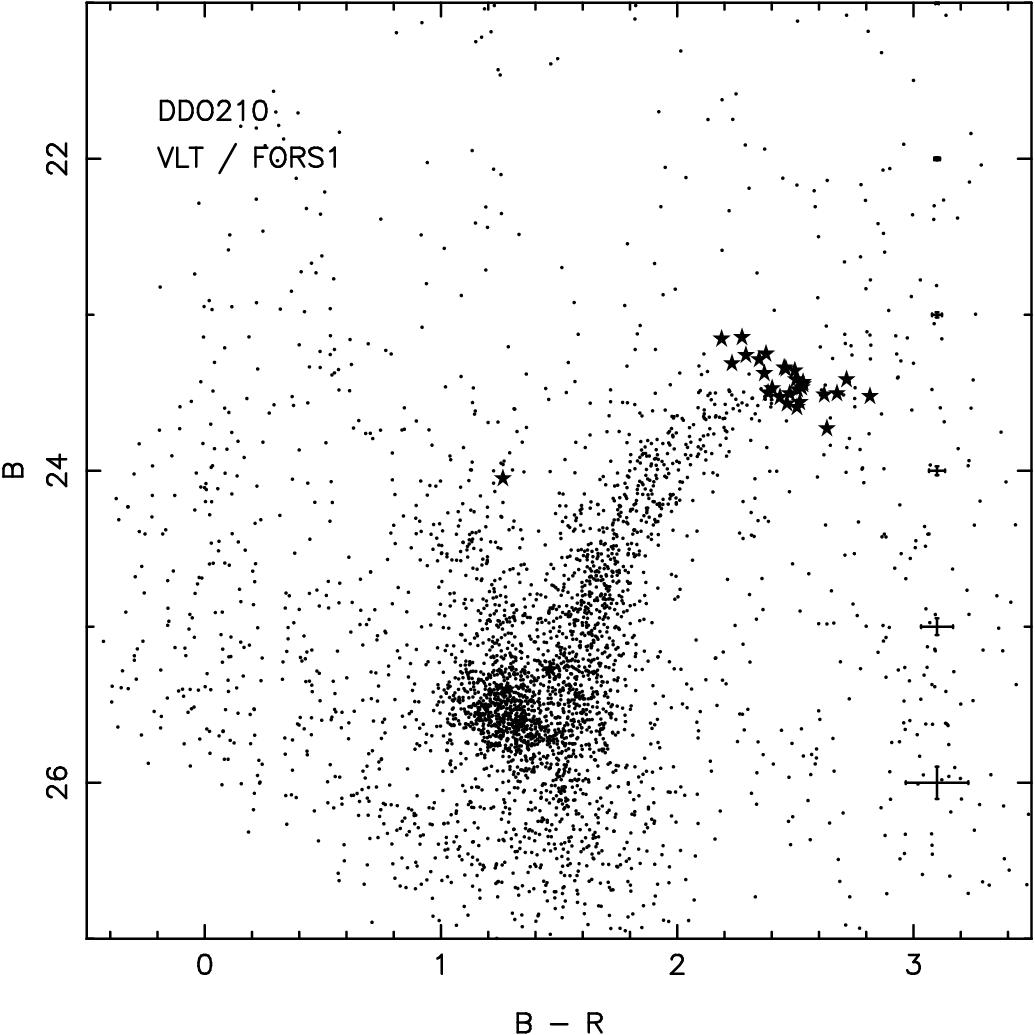}
    \caption{$B$ versus $\left(B - R\right)$ CMD for DDO210, from our
    VLT FORS1 imaging. The highlighted stars have $1.3 < \left(V -
    I\right) < 1.7$ and $20.80 < I < 21.21$, and represent the grouping
    of stars identified by L99 and K02 as marking the TRGB
    (Section 3.1, 3.2). However, the tight grouping of most of these
    stars in this figure slightly to red of the RGB suggest that it is
    far more likely that these stars represent an evolved AGB
    population in DDO210.}
    \label{bcmd}
  \end{center}
\end{figure}

Figures~\ref{vcmd} and \ref{icmd} show the $V$ vs $\left(V - I\right)$
and $I$ vs $\left(V - I\right)$ CMDs for DDO210, respectively,
constructed from the innermost $4$\,arcmins $\times 4$\,arcmins of our
Suprime-Cam field. Also shown are the corresponding CMDs for a
reference area, which lies outside the innermost $7.5$\,arcmins
$\times 7.5$\,arcmins of the Suprime-Cam field, beyond the limits of
DDO210. Figure~\ref{bcmd} shows the $B$ vs $\left(B - R\right)$ CMD
from the FORS1 field. DDO210 is an intrinsically very faint system,
and the CMDs are correspondingly rather sparse. However, several
different stellar populations can be identified in these diagrams,
namely:

\begin{enumerate}
\item{a Galactic foreground sequence. At the brightest magnitudes this
has a colour of $\left(V - I\right) \simeq 0.7$, becoming redder at
fainter magnitudes. This feature is unassociated with DDO210;}
\item{a relatively thin RGB, starting at $I \sim 21$\,mags, with a
colour of $\left(V - I\right) \sim 1.1$\,mags at $V \sim 25$, and
$\left(V - I\right) \sim 1.4$\,mags at $V \sim 23$;}
\item{a red clump of stars, most clearly visible at $B \simeq
  25.8$\,mags and $\left(B - R\right) \simeq 1.2$\,mags;}
\item{bright main sequence stars, at $V \gtrsim 23$\,mags and $\left(V
- I\right) \sim -0.2$\,mags;}
\item{bright blue loop stars, at $V \gtrsim 21$\,mags and $\left(V -
I\right) \sim 0.2$\,mags.}
\end{enumerate}

In this section, we analyse each of these stellar evolutionary phases
in turn, to determine what they are able to reveal about the SFH of
DDO210, in particular regarding the ages and metallicities of the
populations present. In order to do this successfully, we first
revisit the distance estimate to this galaxy.

\subsection{The distance to DDO210 revisited}

\begin{figure*}
  \begin{center}
    \includegraphics[angle=270, width=15cm]{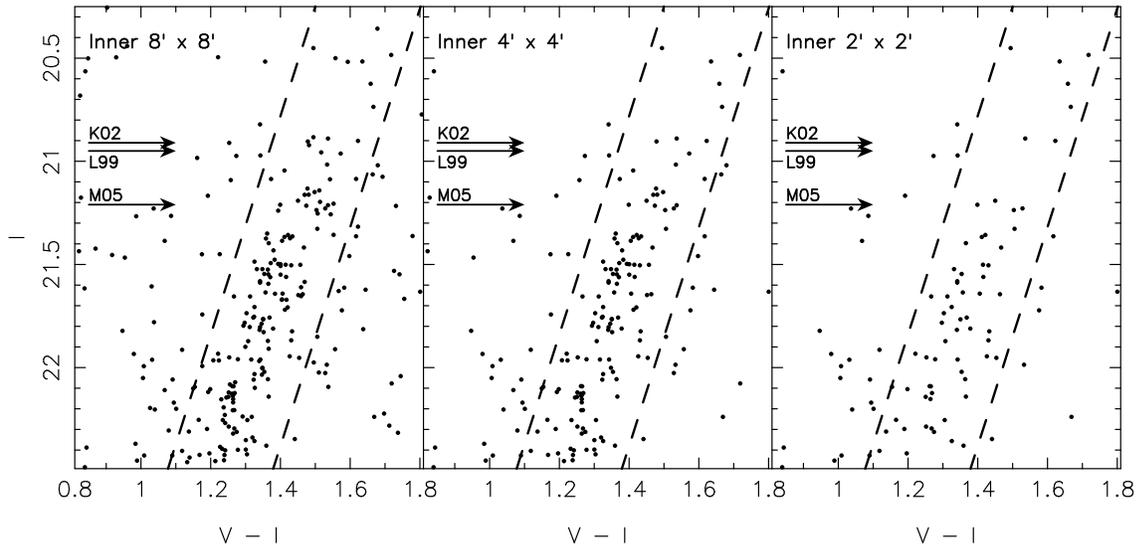}
    \caption{$I-$band CMDs of DDO210 in the region of the TRGB, using
    different spatial cuts on our Suprime-Cam data. Left panel: only
    stars located within the central $4$\,arcmins $\times 4$\,arcmins
    of DDO210; middle panel: only stars located within the central
    $2$\,arcmins $\times 2$\,arcmins; right panel: only stars located
    within the central $1$\,arcmins $\times 1$\,arcmins. The arrows
    indicate the location of the TRGB as derived by Karachentsev et
    al. (2002) (top), Lee et al. (1999) (middle), and McConnachie et
    al. (2005) (bottom). The feature identified by the former two
    authors as the TRGB disappears under the more
    stringent spatial cuts. It is possible that this population
    represents a weak upper AGB population.}
    \label{trgb}
  \end{center}
\end{figure*}

Recent distance estimates to DDO210 have all made use of the
TRGB. Physically, this is the point at which low mass RGB stars
undergo the core Helium flash, which begins their core Helium burning
phase and takes them off the RGB. For metal-poor ([Fe/H]$ <
-0.5$), old ($> 2$\,Gyr) stellar populations, the $I$-band
magnitude of this point has been shown to be a good standard candle
(eg. \citealt{dacosta1990,lee1993a,barker2004}), and is marked by a
discontinuity in the $I$-band luminosity function. This discontinuity
can be found using a variety of methods, such as edge detection
algorithms (\citealt{lee1993a,sakai1996}), or maximum likelihood
techniques (\citealt{mendez2002,mcconnachie2004a}). The reliability of
these methods is greatly increased when used on a well-populated
RGB. As noted by M05, this is not the case for DDO210. These authors
identify two possible locations for the TRGB, favouring their fainter
measurement at $I = 21.21 \pm 0.04$\,mags. L99 and K02 derive a
different value for the TRGB, at $I = 20.95 \pm 0.10$\,mags and $20.91
\pm 0.05$\,mags, respectively. This is closer to the second location
identified by M05 as the possible TRGB of DDO210, at $I \sim
20.8$\,mags.

Although all of the above measurements place DDO210 at $\sim 1$\,Mpc,
the significant disagreement between M05 and L09 + K02 warrants
further investigation. To this end, we examine how the TRGB region of
the $I$-band CMD of DDO210, from our Suprime-Cam data, varies as we
apply different spatial cuts. We expect that the derived value of the
TRGB should remain relatively constant regardless of the spatial cuts
applied so long as the dominant population belongs to DDO210, with the
caveat that cuts which contain fewer stars will be more strongly
affected by small number statistics.  The left panel of
Figure~\ref{trgb} shows an enlargement of the $I-$band CMD from
Figure~\ref{icmd} around the location of the TRGB. The dashed lines
approximately bracket the RGB locus. The arrows indicate the values of
the TRGB as derived by K02 (top), L99 (middle) and M05 (bottom).  The
middle panel shows the same region of the CMD using only stars within
the innermost $4$\,arcmins $\times 4$\,arcmins of the Suprime-Cam
field, and the right panel is the same but using stars within the
innermost $2$\,arcmins $\times 2$\,arcmins. This figure illustrates well
the difficulty in determining the location of the TRGB for an
intrinsically faint system with few bright RGB stars.

Figure~\ref{trgb} shows that the grouping of stars which defines the
L99 and K02 TRGB location has completely disappeared in the smallest
field, and is very weak in the middle field. In contrast, the location
identified by M05 is clear for all of the spatial cuts used, and the
RGB is relatively evenly populated below this point in each
field. Even for a small number of stars, we expect that the RGB
luminosity function will be representatively sampled over its full
magnitude range. It seems most likely, therefore, that the measurement
by M05 provides the most robust estimate of the TRGB of this galaxy,
and implies a distance modulus of $\left(m - M\right)_o = 25.15 \pm
0.08$\,mags ($d = 1071 \pm 39$\,kpc). If this is the case, however,
then the grouping of stars identified by K02 and L99 as the brightest
RGB stars cannot belong to this population. 

We investigate whether the stars immediately brighter than the TRGB
are foreground stars by examining their number density. Using a colour
cut of $1.3 < \left(V - I\right) < 1.7$ and $20.80 < I < 21.21$ to
isolate these stars, we find that there are $25 \pm 5$, $17 \pm 4$ and
$5 \pm 2$ in the $8$\,arcmins $\times 8$\,arcmins, $4$\,arcmins
$\times 4$\,arcmins and $2$\,arcmins $\times 2$\,arcmins fields,
respectively. The uncertainties are Poisson errors. In comparison, a
representative foreground field offset 10\,arcmins from the center of
DDO210 has $5 \pm 2$, $2 \pm 1$ and $0$ stars, respectively, which
satisfy these colour cuts. Thus the number density of these stars in
our DDO210 fields is too large to be explained by a foreground
population alone. 

We believe it likely that these stars represent bright,
intermediate-age, upper asymptotic giant branch (AGB) stars. The stars
occupy the correct locus in the $I$-band CMD to belong to this
population, and the corresponding positions of these stars (satisfying
the same colour cuts as before) in the FORS1 data are highlighted in
Figure~\ref{bcmd}. The majority of these stars are tightly grouped
slightly to the red and brighter than the RGB, again suggesting that
they belong to an evolved AGB population. We show in the next section
that this is consistent with the average age of the stellar
population. We also note that, although the overall number of these
stars is smallest in the smallest field in Figure~8, this is probably
due only to the reduced area. The spatial distribution of these stars
is shown in Figure~12, as open star-symbols overlaid on the grayscale
stellar density map. They are loosely concentrated in the main body of
DDO210, and broadly follow a similar distribution to the overall
stellar population.

\subsection{The candidate asymptotic giant branch population}

In the metal-poor Galactic globular clusters, which consist nearly
exclusively of old stellar populations, no stars are observed brighter
than the TRGB with colours similar to the RGB. This is because the AGB
population terminates at a magnitude comparable to, or less than, the
TRGB. In younger stellar populations, the envelope mass of AGB stars
is larger and so they have the ability to evolve to brighter
luminosities, and can exceed the luminosity of the TRGB. Thus, the
presence of stars at luminosities brighter than the TRGB, with colours
similar to the RGB, is usually interpreted as evidence for the
existence of an intermediate-age population.

The brightest of the candidate AGB stars in DDO210 identified in
Section 3.1 has a magnitude of $I = 20.93$\,mags and a colour of
$\left(V - I\right) = 1.49$\,mags. Correcting for extinction and the
distance to DDO210, this corresponds to an absolute magnitude of $M_I
= -4.32$\,mags and a colour of $\left(V - I\right)_o =
1.42$\,mags. Using the $VI$ bolometric correction derived by
\cite{dacosta1990}, we find that $M_{bol} = -4.86$\,mags for the
brightest candidate AGB star. This number should be considered an
upper limit, due to the difficulties in distinguishing between an
evolved AGB star and a foreground star with a similar colour and
magnitude. A lower limit on the luminosity can be obtained by using
the magnitude of the TRGB; adopting the same colour as before, we find
$M_{bol} = -4.58$\,mags. We therefore adopt $M_{bol} \simeq -4.6 -
-4.9$\,mags for the brightest candidate AGB stars in DDO210.

A recent paper by \cite{rejkuba2006} has a very useful comparison
between the $M_{bol}$ of the brightest AGB star in a selection of
dwarf galaxies with the time since the period of star formation which
produced the star. \cite{mould1979,mould1980} were the first to make
use of such a relation, made possible by the monotonically decreasing
luminosity of AGB stars with age, and this has more recently been
theoretically quantified by \cite{stephens2002}. By comparing DDO210
with those galaxies listed in Rejkuba et al.'s Table~7, we find that
the magnitude of the brightest AGB stars are very similar to those in
Carina and Andromeda~II. For Carina, the last period of star formation
was some 3\,Gyr ago (\citealt{hurleykeller1998}). Figure~19 of Rejkuba
et al. makes explicit their empirical relation between the bolometric
luminosity of the brightest AGB star and the age of the
star. Comparison of our data to this figure reveals that the
bolometric magnitude of the brightest candidate AGB stars in DDO210
imply ages of between $3 - 6$\,Gyr. These numbers should be
interpreted with care, as our analysis is based on very few stars and
we are unable to distinguish between actual AGB stars and
foreground. Nevertheless, these numbers suggest that star formation
was ongoing in DDO210 in the last $3 - 6$\,Gyr.

\subsection{The main sequence and blue loop: evidence of recent star formation}

\begin{figure}
  \begin{center}
    \includegraphics[angle=270, width=8cm]{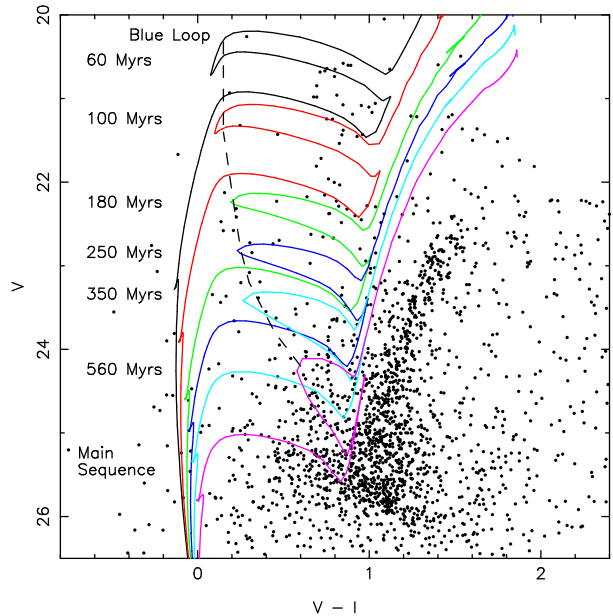}
    \caption{$V$ versus $\left(V - I\right)$ CMD for DDO210, from our
    Subaru Suprime-Cam imaging, with Padova isochrones (Girardi et
    al. 2002) overlaid. These isochrones correspond to [Fe/H] $= -1.3$ and
    ages of 60, 100, 180, 250, 350 and 560\,Myrs. The locus of main sequence
    stars (with $V < 22.8$\,mags, $\left(V - I\right) \simeq -0.2$\,mags)
    and the locus of core Helium burning blue loop stars (with $V <
    20$\,mags, $\left(V - I\right) \simeq 0.2$\,mags) are indicated.}
    \label{young}
  \end{center}
\end{figure}

The presence of a main sequence population and blue-loop stars at
$\left(V - I\right) < 0$ in the CMD of DDO210 reveals the presence of
stars which are significantly younger than those represented by the
evolved AGB stars. Figure~\ref{young} is the $V$ vs $\left(V -
I\right)$ shown in Figure~\ref{vcmd}, with isochrones from
\cite{girardi2002} overlaid. We have shifted these isochrones to the
distance and reddening of DDO210. These isochrones correspond to a
metallicity of [Fe/H] $= -1.3$ (Section 3.5) and ages of 60, 100, 180,
250, 350 and 550\,Myrs. The magnitude of the brightest main sequence
and blue loop stars depends primarily on age, and is relatively
independent of metallicity. They can, therefore, be used as a crude
age indicator.

The dashed line in Figure~\ref{young} marks the approximate locus of
blue loop stars as defined by the isochrones, and it is clear that
they trace the blue loop stars in DDO210 very well. The corresponding
expected locus of red supergiants is, unfortunately, obscured by the
presence of foreground stars with the same colours. The main sequences
of the isochrones match the main sequence population of DDO210 with
notable success. There are a few main sequence stars which are too
blue to match the positions of the isochrones, and it is very possible
that a younger isochrone ($\sim 30$\,Myrs) is required to match this
population. This is what L99 assumed which led them to conclude that
there were stars as young as 30\,Myrs in DDO210. Given the various
uncertainties which go into modelling stellar populations which
inevitably produce variations between different isochrone sets, and
photometric errors, we prefer to conclude that this comparison shows
that DDO210 has stars which formed within the last 60\,Myrs. As noted
in the introduction, this is rather unusual given that star formation
in DDO210 is now observed to be completely dormant
(\citealt{vanzee1997}).

\subsection{The red clump, and limits on an old population}

\subsubsection{The magnitude and colour of the red clump}

\begin{figure}
  \begin{center}
    \includegraphics[angle=270, width=8cm]{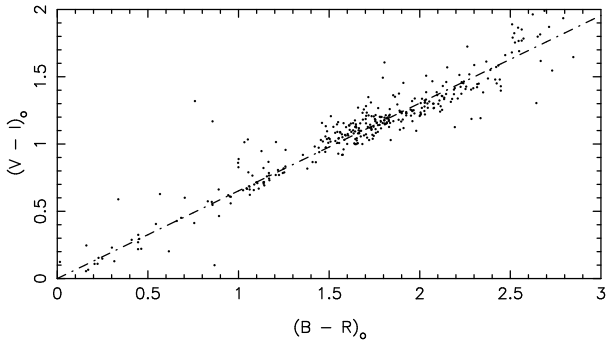}
    \caption{$\left(V - I\right)_o$ vs $\left(B - R\right)_o$ for all
    stars reliably identified in all four filters and with
    uncertainties of $\le 0.05$. The dashed line is the best-fit
    linear relationship of $\left(V - I\right)_o = 0.652 \left(B -
    R\right)_o$.}
    \label{trans}
  \end{center}
\end{figure}

\begin{figure*}
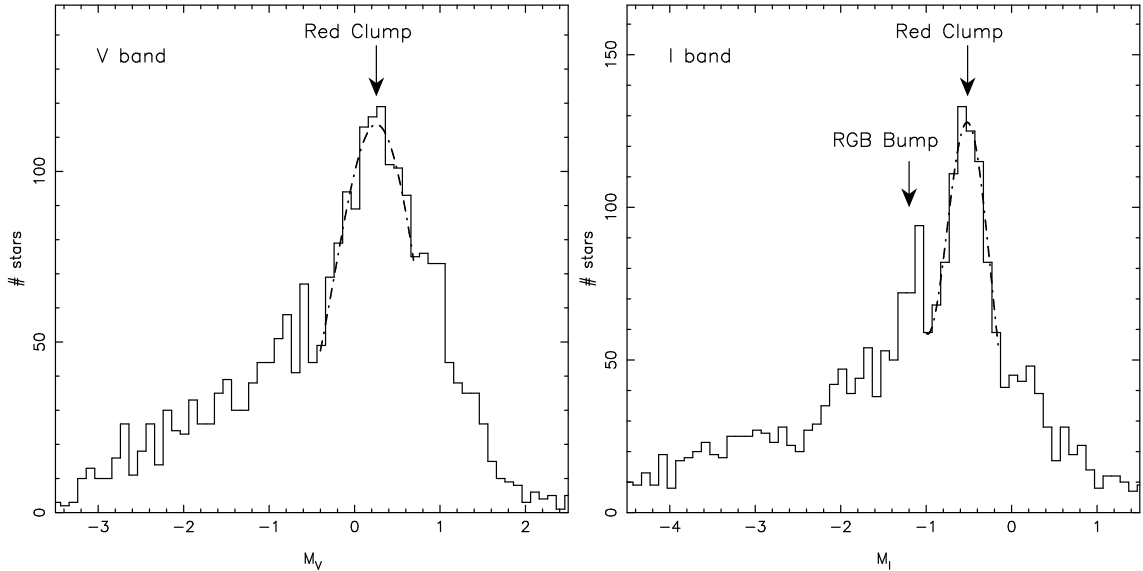

  \begin{center}
    \includegraphics[angle=270, width=7.5cm]{fig8a}
    \includegraphics[angle=270, width=7.5cm]{fig8b}
    \caption{Luminosity functions for $M_V$ and $M_I$ from the
    composite Suprime-Cam/FORS1 data (see text), with a Gaussian fit to
    the region of the red clump overlaid as a dot-dashed line on
    each. The central value of the Gaussian is taken to be the mean
    magnitude of the red clump. The luminosity functions are not
    foreground corrected, but the very high stellar densities in the
    region of interest means that the presence of foreground stars do
    not affect our measurements of the mean magnitudes of the red
    clump.}
    \label{clump}
  \end{center}
\end{figure*}

The most populated region of the CMDs shown in Figures~\ref{vcmd},
\ref{icmd} and \ref{bcmd} is the red clump, which stands out most
clearly in Figure~\ref{bcmd}. Defining the locus of the red clump as
$25.45 < B <26.15$\,mags, $1 < \left(B - R\right) < 1.5$\,mags, we
find that $\sim \frac{1}{4}$ of all stars detected in the FORS1 field
belong to this feature. The red clump is a composite feature in a CMD,
of which the large majority of stars are metal rich, massive, core
Helium burning stars. These are indicative of an intermediate-age
population (2 -- 10\,Gyr). RGB stars and red HB stars can also
contribute to this feature.

A large amount of theoretical modelling of the red clump has been
conducted in recent years
(eg. \citealt{cole1998,girardi1998,girardi2001}), motivated by claims
that the mean magnitude of the red clump can be used as a standard
candle (\citealt{paczynski1998,udalski1998}). This has resulted in a
relatively good understanding of the behaviour of the luminosity and
colour of the red clump on the age and metallicity of the
population. In particular, \cite{girardi2001} calculate the mean
$I$-band magnitude of the red clump, $M_I^{RC}$, as a function of age
and metallicity. The form of this dependency is given in their
Equation~3, and is an integral over a function of the initial mass
function. It transpires that $M_I^{RC}$ is mostly dependent upon age,
whereas the colour of the red clump, $\left(V - I\right)_o^{RC}$, is
mostly dependent on metallicity. We can therefore use the properties
of the red clump to put limits on the SFH of DDO210.

The models of \cite{girardi2001} are formulated in the $V$ and $I$
filters.  As it stands, the Suprime-Cam $I_c$ data easily reaches
to below the red clump, while the $V^\prime$-band exposure is barely
deep enough to fully sample this feature. This therefore leads to some
difficulty in accurately determining $M_I^{RC}$ and $\left(V -
I\right)_o^{RC}$ using this single dataset. The FORS1 data, on the
other hand, is deep enough to fully sample the red clump. In order to
push the $V^\prime$-band data deep enough in order to compare the red
clump properties to these models, we incorporate our deep FORS1 $BR$
photometry into our Suprime-Cam data. To do this, we cross correlate
the two datasets and identify those stars which are reliably
detected in all four filters, and which have small errors in their
photometry ($\le 0.05$\,mags). We then calculate the linear
transformation between $\left(V - I\right)_o$ and $\left(B -
R\right)_o$, shown in Figure~\ref{trans}, and find

\begin{equation}
\left(V - I\right)_o =  0.652 \left(B - R\right)_o~.
\end{equation}

\noindent When combined with Equation~\ref{sub2lan}, this gives

\begin{eqnarray}
M_V & = & M_{I_c} +  0.593 \left(B - R\right)_o\nonumber\\
M_I & = & 1.096 M_{I_c} - 0.097 M_V~.
\label{noV}
\end{eqnarray}

\noindent We use these new transformations to calculate $V$ and $I$
magnitudes for those stars which are detected in the $B$, $R$ and
$I_c$ filters, but not $V^\prime$. The new $V$ vs $\left(V - I\right)$
CMD derived in this way is shown in the left panel of
Figure~\ref{model}, and the red clump is now clearly visible and much
better defined than in Figure~\ref{icmd}.

Figure~\ref{clump} shows the $M_V$ and $M_I$ luminosity functions for
DDO210. The luminosity functions have not had a foreground correction
applied, as for the purposes intended here this makes no practical
difference since the red clump massively outnumbers any foreground
component with the same colours. We measure the mean magnitude of the
red clump by fitting a Gaussian to the luminosity functions in the
region of the red clump, and taking the peak of the Gaussian as the
mean magnitude of the clump.  The resulting fits are shown as
dot-dashed lines in Figure~\ref{clump}. We find that $M_I^{RC} = -0.51
\pm 0.10$\,mags, $M_V^{RC} = 0.27 \pm 0.14$\,mags, and $\left(V -
I\right)_o^{RC} = 0.78 \pm 0.05$. The uncertainties in $M_I^{RC}$ and
$M_V^{RC}$ include the uncertainty in the distance modulus of DDO210.

Figure~1 of \cite{girardi2001} shows the dependency of $M_I^{RC}$ on
age for a range of metallicities. By comparing the value of $M_I^{RC}
= -0.51 \pm 0.10$\,mags for DDO210 with this graph, we find that {\it
the red clump in DDO210 belongs to a population with an average age
equal to or younger than $\sim 6$\,Gyr old} ($4^{+2}_{-1}$\,Gyr). For
a given metallicity, there are as many as three solutions for
$M_I^{RC}$. One of these solutions corresponds to a stellar population
older than $\sim 2$\,Gyr, while the remaining solutions correspond to
populations younger than $\sim 2$\,Gyr. The average metallicity of the
dominant population, as implied by $\left(V - I\right)_o^{RC} = 0.78
\pm 0.05$, is $-2.0 \lesssim {\rm [Fe/H]} \lesssim -1.3$, where the
lower limit is least well constrained.

\subsubsection{The shape of the red clump}

\begin{figure*}
  \begin{center}
    \includegraphics[angle=270, width=14.cm]{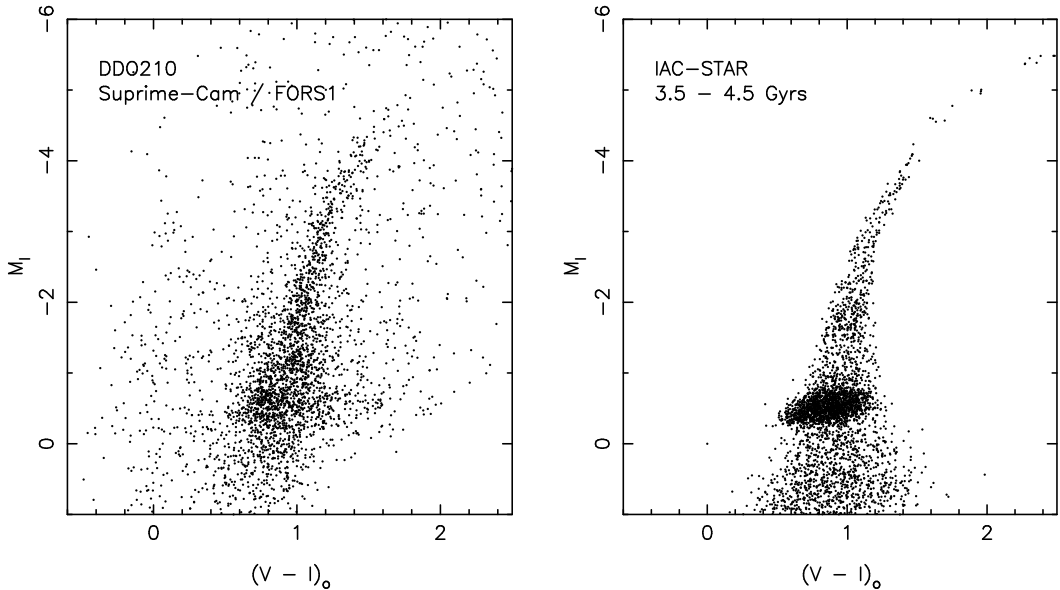}
    \caption{Left panel: $M_I$ vs $\left(V - I\right)_o$ CMD using our
    composite Suprime-Cam/FORS1 dataset. The red clump is now clearly
    visible. Right panel; Illustration of a partial SFH model
    generated with IAC-STAR (Aparicio \& Gallart 2004), scaled for the
    same number of observed stars, and subsequently convolved with our
    error distribution. This CMD is for the stellar population which
    formed its stars at a constant rate between $3.5 - 4.5$\,Gyr ago,
    and which reproduces the mean magnitude of the red clump very
    well.}
    \label{model}
  \end{center}
\end{figure*}

\begin{figure*}
  \begin{center}
    \includegraphics[angle=270, width=12.cm]{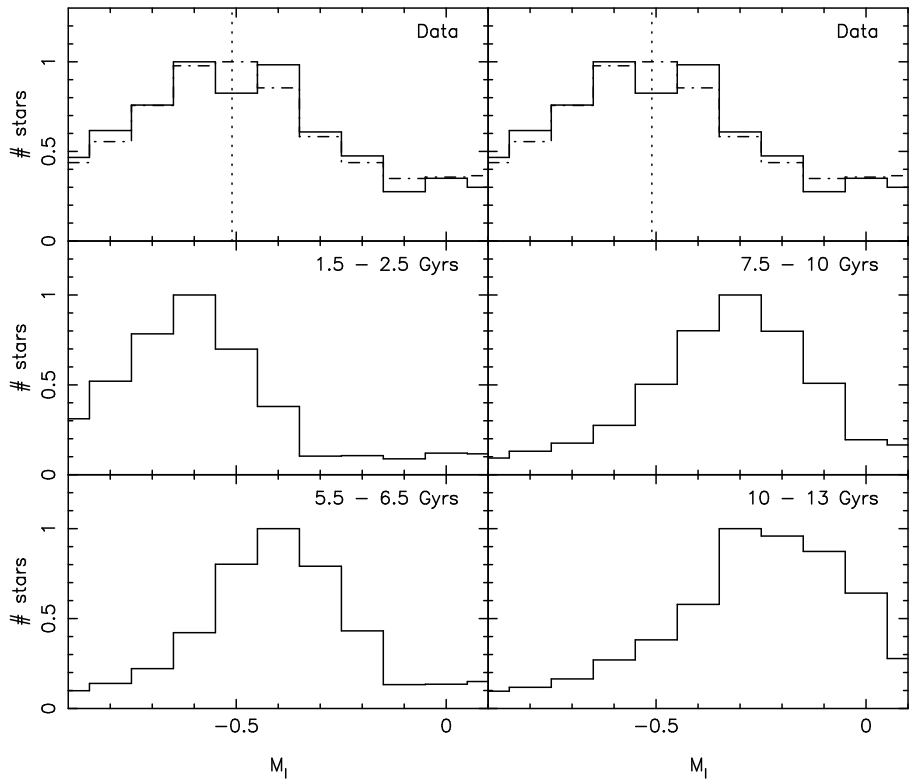}
    \caption{Top panels: $M_I$ luminosity function for the red clump
    from our composite Suprime-Cam/FORS1 data. The dot-dashed
    histograms represent a composite stellar population with a constant
    SFR between $1.5 - 3.5$\,Gyr and $4.5 - 6.5$\,Gyr. Other panels:
    $M_I$ luminosity functions for red clumps generated with the
    IAC-STAR tool, convolved with our error distributions. Star
    formation in these models occurred at a constant rate for the
    periods indicated, with the metallicity held constant at the value
    which reproduces the colour of the RGB for those ages.  The young
    -- intermediate age red clumps have peak magnitudes consistent
    with the broad peak of the observed luminosity function. On the
    other hand, the older populations produce red clumps which are too
    faint in comparison to the data. If red clump stars of these ages
    are present in the data, they must be a minority population. No
    individual model reproduces the observed red clump luminosity
    function.}
    \label{rclf}
  \end{center}
\end{figure*}

The {\it magnitude} of the red clump implies that the majority of
stars in DDO210 belong to an intermediate-age population, and so far
we have found no evidence of an old population in this galaxy. To try
to better constrain the range of ages of stars present in DDO210, we
now compare the {\it shape} of the red clump to model CMDs with a
range of SFHs. We use the web-based synthetic CMD tool
IAC-STAR\footnote{http://iac-star.iac.es/iac-star}, developed by
\cite{aparicio2004}. The user supplies the SFH, chemical enrichment
law, initial mass function (IMF) and binary fraction (and mass ratio)
for the stellar population they want to model, and has a choice of
stellar evolution libraries and bolometric correction libraries which
can be used. The program then does a Monte-Carlo simulation of the
evolution of $N$ stars, with a distribution of properties chosen to be
consistent with the user-supplied input, and outputs the luminosities
of these stars at the present day in several filter systems.

The modelling of CMDs is prone to many uncertainties and degeneracies,
such as the evolutionary tracks used, effects of mass loss, binarity,
age-metallicity degeneracies, the shape of the IMF, etc. Excellent
examples of this type of work and reviews of these difficulties can be
found in, among others,
\cite{tolstoy1996,aparicio1997,tolstoy1998,gallart1999a,gallart1999b,aparicio2001,aparicio2004,gallart2005}. These
difficulties are especially problematic when the data does not extend
to beneath the main-sequence turn-offs (which offer the best handle on
the age of the population). Therefore, we keep our analysis
intentionally simple and use it only to emphasise that DDO210 cannot
contain a majority old population. A more sophisticated analysis is
not warranted by the current data given the uncertainties involved.

We use the \cite{bertelli1994} evolutionary tracks and the bolometric
correction libraries of \cite{lejeune1997}. We adopt a power law IMF,
$\phi\left(M\right) \propto M^{-x}$, with an exponent of $x = -1.35,
-2.2$ and $-2.7$ in the range $M \in [0.1,0.5], [0.5,1]$ and
$[1,120]$, respectively. We ignore the effects of binary stars. We
construct 10 different SFHs (`partial models'), corresponding to a
constant SFR between [0,0.5],[0.5,1.5],
[1.5,2.5],[2.5,3.5],[3.5,4.5],[4.5,5.5],[5.5,6.5],[6.5,7.5],[7.5,10]
and [10,13]\,Gyr ago, with no star formation at all other times. For
each interval, the metallicity (chemical enrichment law) is chosen to
reproduce the colour of the RGB (Section~3.5). The magnitudes of the
model stars are then convolved with the error distribution of our data
to facilitate a comparison between the two. The left panel of
Figure~\ref{model} shows a CMD of our composite Suprime-Cam/FORS1
dataset. The right panel is an illustration of an IAC-STAR model, and
shows the Monte-Carlo simulation of the $3.5 - 4.5$\,Gyr stellar
population, normalised to the same number of stars and convolved with
our error distribution. We note that a bright ($M_I< -4$\,mags), AGB
population is produced by this model. This resembles the population of
stars just brighter than the TRGB in our data which were discussed at
length in Sections 3.1 and 3.2.

The solid histograms in the top panels of Figure~\ref{rclf} shows the
normalised $M_I$ luminosity function for DDO210 in the region of the
red clump. The other 4 panels show the red clump luminosity function
for 4 of our simulated partial SFHs. The youngest ($1.5 - 2.5$\,Gyr)
red clump peaks at a brighter magnitude than the data, although this
young population could contribute to the bright side of the DDO210 red
clump. Similarly, the $5.5 - 6.5$\,Gyr population could contribute to
the faint side of the red clump. A single, $1$\,Gyr, period of star
formation at intermediate ages does not produce a red clump with the
required magnitude dispersion to match the data. To illustrate this,
the dot-dashed histograms overlaid on the top panels of
Figure~\ref{rclf} show the red clump luminosity function for a star
formation history which is constant between $1.5 - 3.5$\,Gyr and $4.5
- 6.5$\,Gyr, and zero at all other times. This broader, composite SFH
reproduces both the {\it magnitude} and the {\it shape} of the red
clump.

Older stellar populations are shown in the remaining two panels of
Figure~\ref{rclf}. The $7.5 - 10$\,Gyr red clump consists
predominantly of stars fainter than are seen in the red clump of
DDO210. For stars older than $10$\,Gyr, the region of the CMD which
contains the red clump contains a lot of HB stars. The luminosity
function in this region is significantly fainter than the observed
luminosity function. It is clear that, if stars older than $\sim
7$\,Gyr are found in DDO210, then their overall contribution to the
stellar population of DDO210 must be $<<50\%$. Otherwise, the
morphology of the red clump in DDO210 would be notably different. We
estimate that, if such a population exists, it probably contributes no
more than $\sim 20 - 30\%$. This estimate is unaffected by
incompleteness, since the red clump is above our incompleteness
limit. However, deeper photometry, which sample some of the older main
sequence turn-offs in DDO210, would be of huge value in providing much
stronger limits on the contribution of an ancient stellar population
in DDO210.

\subsection{The red giant branch and its bump}

\begin{figure*}
  \begin{center}
    \includegraphics[angle=270, width=12cm]{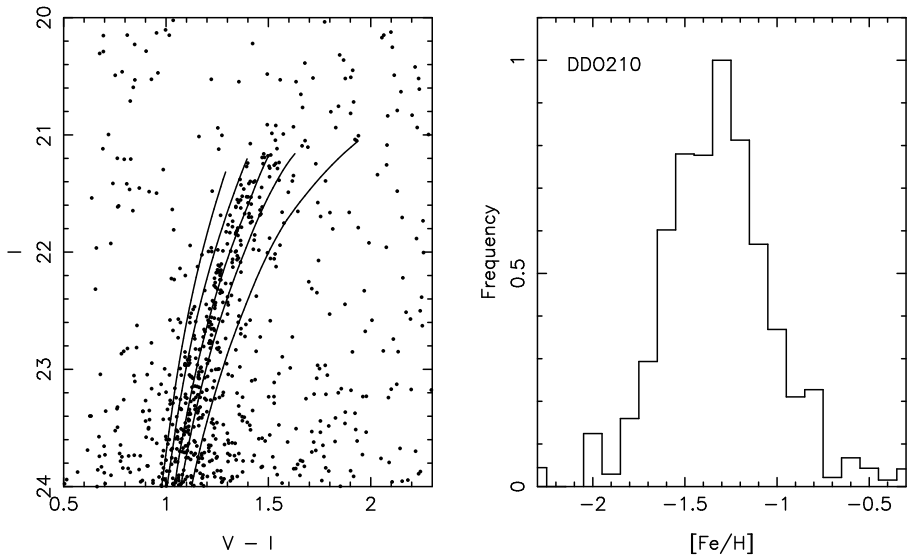}
    \caption{Left panel: the upper RGB of DDO210. Victoria -- Regina
    isochrones by VandenBerg et al. (2005), with $BVRI$ colour --
    $T_{eff}$ relations as described by VandenBerg \& Clem (2003), are
    overlaid. These tracks correspond to an age of 4\,Gyr and a
    metallicity of [Fe/H] $= -2.3, -1.8, -1.4, -1.1$ and $
    -0.8$. These are a few of the larger grid of isochrones which
    we use to construct the MDF. Right panel: MDF for DDO210, derived
    by interpolation of the colours and magnitudes of the RGB stars in
    DDO210 with the full grid of Victoria-Regina isochrones
    corresponding to the 4\,Gyr population. The mean metallicity found
    in this way is [Fe/H] $= -1.3 \pm 0.1$, with a dispersion of
    $\sigma_{{\rm [Fe/H]}} = 0.27$\,dex. }
    \label{mdf}
  \end{center}
\end{figure*}

The most prominent feature in the CMDs displayed in
Figures~\ref{vcmd}, \ref{icmd} and \ref{bcmd} is the RGB. This feature
is only seen in stellar populations older than a few Gyr, and is an
indicator of an intermediate and/or old population. It cannot be used
as a more precise indicator than this without certain difficulties,
because of the well known degeneracies of RGB colour with both age and
metallicity (redder RGB stars may be older and/or more metal rich than
bluer RGB stars). When analysing a CMD, the standard assumption which
is used in the absence of other constraints is that the stellar
populations which give rise to the RGB are as old as the MW globular
clusters. In this case, the colour and colour dispersion of the RGB
are due to metallicity variations, and so an estimate of the
metallicity may be made. This is usually obtained by a bilinear
interpolation of the position of the RGB stars in colour -- magnitude
space between a set of globular cluster fiducial tracks, isochrones or
evolutionary tracks, and creating a metallicity distribution function
(MDF; eg. \citealt{durrell2001}). Alternatively, empirical
calibrations between RGB colour and metallicity exist which can be
easily applied (eg. \citealt{dacosta1990}).

L99 assume that DDO210 is as old as the MW globular clusters, and
derive a mean metallicity of [Fe/H] $= -1.9 \pm 0.1$, where the second
term is the error in the mean metallicity, not the metallicity
dispersion. Using our Suprime-Cam data, we construct a MDF for DDO210
by interpolating between the Victoria - Regina set of isochrones by
\cite{vandenberg2005} for a 13\,Gyr stellar population, with $BVRI$
colour-$T_{eff}$ relations as described by \cite{vandenberg2003}. We
use only those stars in the brightest 1.5\,mags of the RGB, where the
photometric errors are smallest and where the colour difference caused
by metallicity variations is largest. We find that, under this
assumption, the mean metallicity is [Fe/H] $= -1.9$, in agreement with
L99. However, from Section~3.4, we know that the majority of stars in
DDO210 do not belong to an old population, and that most of the stars
have an average age of $\sim 4$\,Gyr. We therefore repeat our
analysis, but this time compare the RGB stars to isochrones which are
4\,Gyr old. The resulting MDF can be seen in the right panel of
Figure~\ref{mdf}. The left panel shows a zoomed in view of the RGB of
DDO210, with a few of the isochrones which are used overlaid. With
this assumption about the age, we find DDO210 has a mean metallicity
of [Fe/H] $= -1.3 \pm 0.1$, and a metallicity dispersion of
$\sigma_{{\rm [Fe/H]}} = 0.27$\,dex. The latter quantity is likely
overestimated, however, since some of the colour variation in the RGB
will be due to the age spread which we know exists in DDO210
(Section~3.4.2).

Finally, the RGB bump is clearly visible in the $I$-band luminosity
function shown in Figure~\ref{clump} at $M_I \simeq -1.2$\,mags. It is
less obvious in the $V$-band luminosity function, where its magnitude
is closer to that of the red clump. However, from the CMD in
Figure~\ref{model}, its colour is $\left(V - I\right) \simeq 1$, and
so $M_V \simeq -0.2$\,mags. The RGB bump was first discovered in dwarf
galaxies by \cite{majewski1999b} in Sculptor (see also
\citealt{bellazzini2001,bellazzini2005,monaco2002}). As the helium
burning core of the RGB star increases in size, so too does the radius
of the outer hydrogen-burning shell. The RGB bump represents the point
at which this shell encounters the chemical discontinuity left behind
by the innermost penetration of the convective envelope. This causes a
decrease in luminosity while the hydrogen-burning shell adapts to this
new environment, before the star continues its ascent of the RGB. The
magnitude of the RGB bump and its dependence on age and metallicity
has been quantified for MW globular clusters by
\cite{ferraro1999}. Using their Equation~3, the RGB bump is expected
at $M_V = -0.27$ for a population with an age of $4$\,Gyr and a
metallicity of [Fe/H] $= -1.3$. This compares favorably to where we
detect the bump in DDO210. To the best of our knowledge, this is the
first time that the RGB bump has been observed in a dwarf galaxy
beyond the MW satellite system.

\section{The structural characteristics of DDO210}

Our photometric datasets extend over a relatively large area of sky,
and the Suprime-Cam data in particular is well suited to studying the
global structural properties of DDO210, both in terms of the
unresolved light and the resolved stellar components. For example,
having shown that DDO210 consists of composite stellar populations, a
natural question to ask is {\it how are these populations spatially
distributed?} In this section, we first derive some global parameters
for DDO210 based upon both its resolved and unresolved components. We
then study the distribution of the various stellar types in DDO210,
and finally we compare its stellar structure to its HI distribution,
based upon VLA observations by \cite{young2003}.

\subsection{The integrated structure}

Following the method of \cite{mcconnachie2006b}, we have directly
estimated the central surface brightness and integrated luminosity of
DDO210 from the integrated $V^\prime$ flux distribution. From
Equation~\ref{sub2lan}, the colour correction to convert to Landolt
$V$ is significantly smaller than the final uncertainties in our
measurements, and so we can treat our measurements as Landolt
$V$. The object catalogues are used to define a bright
foreground star component 1\,magnitude above the TRGB, to allow for the
presence of AGB stars. A circular aperture is excised around each
foreground star and the flux within this aperture is set to the local
sky level, interpolated from a whole-frame background map. The size of
this aperture is the maximum of four times the catalogue-recorded area
of the bright star at the detection isophote, or a diameter four times
the derived FWHM seeing. Each frame is then re-binned on a $4 \times
4$ grid to effectively create $0.8$\,arcsec pixels. The binned image
is then further smoothed using a 2D Gaussian filter with a FWHM of
5\,arcsecs.

The result of this procedure is to produce a coarsely sampled smooth
image containing both the resolved and unresolved light contribution
from DDO210.  The central surface brightness can then be measured by
deriving the radial profile, here defined as the median flux value
within elliptical annuli.  Finally, large elliptical apertures are
placed over DDO210 and several comparison regions to estimate the
background-corrected integrated flux from the dwarf and the reference
regions.  The variation in the flux from the multiple comparison
measures gives a good indication of the flux error, which is, of
course, dominated by systematic fluctuations rather than by random
noise.  This is a particular problem for DDO210 because of the
numerous resolved blue stars we observe in this galaxy. To mitigate
the effect of random residual foreground stellar halos and scattered
light from bright stars just outside the field of view, the elliptical
apertures are chosen to correspond to the derived value of
$r_{\frac{1}{2}} = 1.1$\,arcmins (Section~4.2). The estimated total
flux is then scaled to allow for this correction. The small angular
size of DDO210 also allows us to measure the integrated flux by
summing over the whole galaxy, which we can then compare with the
value derived from within $r_{\frac{1}{2}}$.

Using the flux within $r_{\frac{1}{2}}$, we find that the total
magnitude of DDO210 is $V = 14.74 \pm 0.10$\,mags, which implies $M_V
= -10.58 \pm 0.13$\,mags. The total flux obtained on integrating over
the entire galaxy is, for comparison, $V = 14.49 \pm 0.10$\,mags,
implying $M_V = -10.83 \pm 0.13$\,mags. We believe that the
discrepancy between these two values is a result of the flux in $V$
being centrally concentrated due to a central population of young stars
(Section~4.2). The extinction-corrected central surface brightness of
DDO210 is $\mu_o = 23.6 \pm 0.2$\,mags per sq.\,arcsec. These values
confirm previous results showing that DDO210 is one of the faintest
galaxies in the Local Group.

\subsection{The stellar structure}

\begin{figure*}
  \begin{center}
    \includegraphics[angle=270, width=8cm]{fig12a}
    \includegraphics[angle=270, width=8cm]{fig12b}
    \includegraphics[angle=0, width=7.25cm]{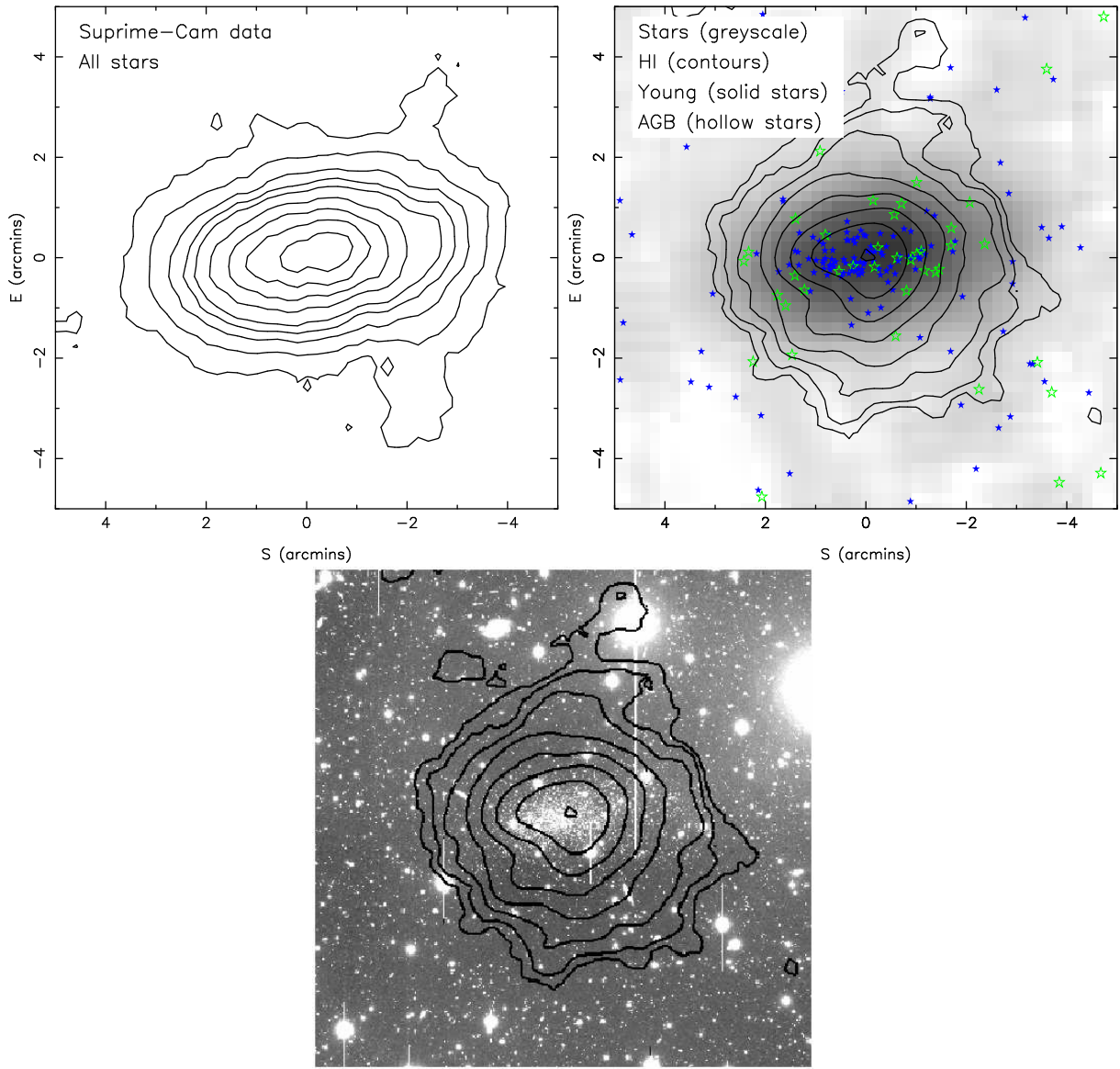}
    \caption{Top left panel: a contour map of the spatial distribution
    of all stellar sources in DDO210 which have $I_c <
    26$\,mags. Contour levels are set at 2, 5, 10, 15, 20, 30, 40, 50
    and 60\,$\sigma$ above the background. Top right panel: HI
    contours for DDO210, from low resolution VLA observations by Young
    et al. (2003). Contour levels are set at $0.1, 0.2, 0.4, 0.8, 1.6,
    3.2, 6.4$ and $12.8 \times 10^{20}$\,atoms\,cm$^{-2}$. The stellar
    distribution shown in the left panel is included as a linear
    grayscale. The solid stars represent the positions of candidate
    main sequence and blue loop satisfying $I < 24.3$\,mags and
    $\left(V - I\right) < 0.5$, and the open stars represent the
    positions of candidate bright AGB stars, satisfying $1.3 < \left(V
    - I\right) < 1.7$ and $20.80 < I < 21.21$. Bottom panel: The
    $I_c$-band Suprime-Cam image of DDO210, with HI contours
    overlaid. The scale, orientation and contour levels are the same
    as in the previous panels.}
    \label{cont}
  \end{center}
\end{figure*}

Figure~\ref{image} shows the $V$ and $I$-band Suprime-Cam images of
DDO210. Although the galaxy is clearly visible at the center of each
field, it is hard to tell from these images how far out its stars
extend. In order to probe to fainter surface brightness thresholds,
therefore, we construct a contour plot of the resolved stars. This
approach can probe several magnitudes fainter in surface brightness
than studies based only on the unresolved light (typically $\mu >
31$\,mags per sq.\,arcsec$^{-1}$).

We construct our contour map following the methodology of
\cite{irwin1995} and \cite{mcconnachie2006b}. Briefly, we use all
those objects which have been reliably classified as stellar in the
$I$-band of the Suprime-Cam data, and which have $I_c <
26$\,mags. We construct an image from this data by binning stars
according to their position in a grid with a resolution of
12\,arcsecs. We then apply a crowding correction to this grid
(\citealt{irwin1984}), such that

\begin{equation}
f^\prime \simeq f \left( 1 + 2 f A^\prime +\frac{16}{3} f^2 A^{\prime 2} ... \right)~.
\label{crowd}
\end{equation}

\noindent where $f$ is the observed number density of {\it all
detected images} (not just stellar), $f^\prime$ is the actual number
density and $A^\prime$ is the typical area of the image. Adopting this
correction increases the number density of stars in the very central
region by a factor of $\sim 1.7$, but has little effect outside the
inner $\sim 1$\,arcmin. 

The top left panel of Figure~\ref{cont} shows the contour map of the
spatial distribution of the stellar sources in DDO210. A linear filter
has been applied to this map to smooth out structures below a scale of
1.4\,arcmins. The contour levels have been set at $2, 5, 10, 15, 20,
30, 40, 50$ and $60\,\sigma$ above the background. In order to
estimate the background level, we constructed a histogram of the
intensity distribution of pixels located in the outer parts of the
Suprime-Cam field. We then performed a sigma-clipped fit of a Gaussian
to this distribution, finding that the background stellar intensity is
$32.57 \pm 0.14$\,stars\,arcmin$^{-2}$, and $\sigma = 7.71 \pm
0.12$\,stars\,arcmin$^{-2}$.

DDO210 has a fairly regular overall stellar structure for a dwarf
galaxy, with no obvious substructures or irregular shaped
isophotes. Its position angle and ellipticity can be quantified as a
function of isophotal threshold in an identical way as in
\cite{mcconnachie2006b}, making use of standard image-analysis
routines to calculate these quantities in a parameter-independent
fashion. The position angle of DDO210, measured east from north, is
$\theta = 98\degg7 \pm 1\degg0$, where the uncertainty indicates the
variation in this quantity for the different isophotal
thresholds. Clearly, the variation is minimal and no trends, such as
isophotal rotation, are observed. The ellipticity ($\epsilon = 1 -
b/a$), on the other hand, varies systematically as a function of
radius in DDO210, such that the galaxy is more elongated at smaller
radius ($\epsilon \sim 0.6$) and more circular at larger radius
($\epsilon \sim 0.4$). While this effect is not dramatic, it is
sufficiently strong that it can be detected by eye in the top left
panel of Figure~\ref{cont}.

\begin{figure}
  \begin{center}
    \includegraphics[angle=270, width=8cm]{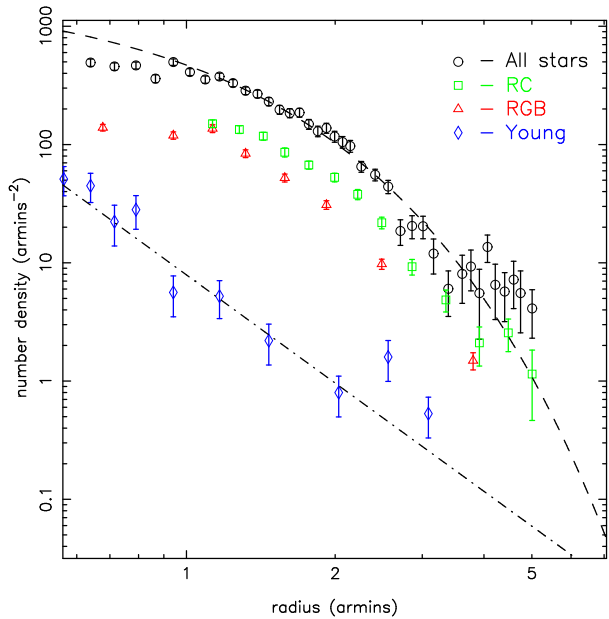}
    \caption{Radial profiles of various stellar populations in DDO210:
    all stars (black circles), red clump (green squares), RGB stars
    (red triangles) and young stars (blue diamonds). See text for
    colour cuts used. The RGB stars and red clump stars follow a
    similar exponential profile to the overall population (dashed
    curve). The young stars, however, follow a different radial
    profile, represented here by a power-law (dot-dashed
    line). Crowding may affect these profiles inside $\sim
    0.8$\,arcmin.}
    \label{profile}
  \end{center}
\end{figure}

The spatial distributions of the various stellar populations
identified earlier can also be investigated separately by applying colour
cuts on the CMD to isolate the stars belonging to these features. In
particular, we isolate the following populations;

\begin{enumerate}
\item{RGB stars. These are defined as stellar objects with $21.21 < I
  < 23.23$\,mags, $I > -5.3 \times \left(V -I\right) + 28.2$, $I <
  -5.3 \times \left(V -I\right) + 29.8$;}
\item{red clump stars.  These are defined as stellar objects with
$24.3 < I_c < 25.2$\,mags;}
\item{main sequence and blue-loop stars. These are defined as stellar
objects with $I <24.3$\,mags, $\left(V - I\right) < 0.5$\,mags.}
\end{enumerate}

Radial profiles are constructed for each of these populations by
counting the number of stars in elliptical annuli, centered on DDO210,
with a fixed position angle of $\theta = 98\degg7$ and a fixed
ellipticity of $\epsilon = 0.5$. Using fixed values for these
parameters ensures that the radial profile derived will be robust and
suitable for comparison to models. Crowding and background corrections
are applied in a similar way as for the contour plot. Results are
shown in Figure~\ref{profile}. Green squares represent the red clump
component, red triangles represent the RGB component, and blue
diamonds represent the young stellar component. Also shown as black
circles is the radial distribution of the overall stellar population,
defined in the same way as for the contour plot. Error bars show the
combined uncertainty of the Poisson error in the counts and the
uncertainty in the background estimate. Each bin has a minimum
signal-to-noise ratio of 10, with the exception of the young stellar
profile. Here, the fewer stars available require us to use a lower
signal-to-noise threshold of 2.5. The radial profiles stop at the
point at which they level out, indicating that we can no longer trace
the DDO210 population. The inner $\sim 0\degg8$ of some of the
profiles may be unreliable due to crowding effects, but the outer
parts of the profile remain unaffected.

The dashed curve in Figure~\ref{profile} is an exponential fit to the
overall stellar population of DDO210, and has a scale-length of $r_e =
0.66 \pm 0.02$\,arcmins (half light radius, $r_{\frac{1}{2}} = 1.1 \pm
0.03$\,arcmins), corresponding to $r_e = 206 \pm 10$\,pc at the
distance of DDO210. Only points beyond $1$\,arcmin were used in the
fit to avoid crowding issues. An exponential function with a similar
scale length also matches the RGB and red clump radial profiles
equally well ($\chi_{reduced}^2 = 1.8$). This is unsurprising, given
that most of the stars in DDO210 are red clump or RGB stars, and that
it is probable that most of these stars arise in the same star
formation episode. 

On the other hand, the youngest stars in DDO210 are not distributed in
the same way as the older populations. The radial distribution of the
main-sequence and blue-loop stars is better approximated by a power
law (dot-dashed line), of index $n = -3.0 \pm 0.3$ ($\chi_{reduced}^2
= 1.5$). Clearly, the recent star formation in this galaxy occurred in
a spatially more concentrated region than the earlier epoch(s) of star
formation. This difference in the radial profiles of the young and
older stellar populations confirms and quantifies the suggestion by
L99 that the central regions of DDO210 have seen an enhancement in the
SFR during the last few hundred Myrs.

The radial profiles of the older stellar populations in DDO210 bears
some similarity to the radial profile of Leo~A, derived by
\cite{vansevicius2004}. These authors found that, beyond a radius of
$\simeq 5.5$\,arcmins, the slope of the radial profile of Leo A
changes and becomes shallower (their Figure~3). They argued that this
represented a detection of a halo component in Leo~A. DDO210 is
sufficiently faint that we are unable to trace the stellar component
out to much beyond $\sim 5$\,arcmins, to see if there is a similar
change in slope for this galaxy. The distribution of younger stars in
Leo~A is also different to its older populations, but in Leo~A the
gaseous distribution is similar to the RGB stars. As we show in the
next section, this is not the case for DDO210.

\subsection{The gaseous structure}

Given our findings in the previous section, it is interesting to
explore the spatial distribution of the gas in DDO210, to see if and
how this correlates with the stellar distribution, especially the
young stars. The HI content of DDO210 has been relatively
well-studied, starting with the survey by \cite{fisher1975}. More
recently, \cite{lo1993} have measured the HI mass in DDO210 to be $3
\times 10^6\,M_\odot$. \cite{young2003} measure the radius to which
the HI column density has dropped to $10^{20}$\,atoms\,cm$^{-2}$, and
compare this to the radius of the $25$\,mags per sq\,arcsec stellar
isophote. They find that the ratio of these HI to optical scale
lengths is approximately 2:1, which is typical of dIrr galaxies. The
peak column density of HI is likewise typical, at $\sim 1.3 \times
10^{21}$\,atoms\,cm$^{-2}$. L.~Young has kindly sent us her HI data
for DDO210, and we have plotted the low-resolution data onto a
grayscale image of the stellar distribution (linear scale) in the top
right panel of Figure~\ref{cont}. Also plotted as individual points in
this panel are the candidate main sequence and blue plume stars from
our Suprime-Cam imaging. The bottom panel of Figure~\ref{cont} shows,
on the same scale, the HI contours overlaid on the $I$-band
Suprime-Cam image of DDO210.

It is clear from Figure~\ref{cont} that the overall stellar and HI
spatial morphologies of DDO210 are different, insofar as the HI
distribution is predominantly spherical, whereas the stellar
distribution is much more elliptical. \cite{young2003} noted that
DDO210 displayed a large scale velocity gradient, and later
observations by \cite{begum2004} measured a rotation curve for the HI,
which peaks at $\sim 16$\,km\,s$^{-1}$. The kinematic major axis of
the rotation shows significant warping, and its position angle changes
substantially with radius. However, this variation generally follows
the morphological position angle derived from fits to the HI
distribution. At $\sim 2$\,arcmins from the center of the HI
distribution, the position angles of the kinematic and morphological
HI axes are $\sim 95^o$, which is in good agreement with the position
angle we derive for the stellar distribution. \cite{begum2004} also
derive an inclination for DDO210 of $27^o \pm 7^o$, based upon the
assumption that the actual shape of the HI is circular. Using a
similar assumption, we find that the inclination of DDO210 from its
stellar distribution varies between $\sim 55^\circ$ and $\sim
65^\circ$, in agreement with L99 and significantly different to the HI
value. Based on these numbers, we consider it unlikely that the
majority of stars in DDO210 are distributed in a disk, and so the
inclination we derive is unlikely to be meaningful. \cite{begum2004}
reach the same conclusion. We note that the fact that most of the
stars are radially distributed in an exponential profile is {\it not}
evidence that the stars are distributed in a disk. Although
exponential profiles match disks very well, our experience with dSph
galaxies shows that these too are well described by exponential
profiles (\citealt{irwin1995,mcconnachie2006b}; see also
\citealt{faber1983}). We also note that these findings do not prevent
{\it some} stars in DDO210 from being distributed in a disk.

\cite{young2003} and \cite{begum2004} both overlay their HI contours
on optical images of DDO210 and note that the optical component
appears to be offset from the HI component. A comparison of the panels
of Figure~\ref{cont} reveals the cause of this apparent offset. The
lower panel shows that the brightest optical emission and the HI
contours are not centered on one another, whereas the top two panels
show that the overall stellar and HI content of DDO210 are centered on
one another. The reason for this disparity is that the {\it young}
stellar component, to which the brightest stars in DDO210 belong, is
not clumped in the middle of DDO210, but is instead offset a few
arcmins to the east. This is easily visible from the top right
panel. Since a few young stars contribute far more light than many
older stars, the optical emission is biased towards tracing this
population, rather than the majority older population. It is
interesting to note that the highest density of young stars occurs at
a point where there is a slight `dent' in the HI contours, perhaps
suggesting that some event caused the gas to be asymmetrically
compressed, triggering off-center star formation. Alternatively, the
off-center star formation may be due to recently captured gas, which
has not become fully virialised. Finally, we also note that the
stellar component of DDO210 is significantly more extended that
suggested by the optical image, and effectively extends out as far as
the HI.

\section{Discussion: Is this the dawning of the age of Aquarius?}

DDO210 is dominated by an intermediate age stellar population. In this
respect, it resembles Leo~A, another relatively isolated Local Group
galaxy which has been shown to consist of a predominantly young ($<
2$\,Gyr) stellar population (\citealt{tolstoy1998}). Leo~A is located
$\sim 690\,$kpc from the MW (\citealt{mateo1998a}), and it is
conceivable that it is on a highly radial orbit around the
MW. Interactions may have affected its SFH. In particular, the
presence of RR~Lyrae stars in Leo~A shows that it has an old
component, and may well have had ongoing, low-level, star formation
until a few Gyr ago, when the star formation rate increased. This may
have been brought on by a close encounter with the MW which triggered
enhanced star formation.

DDO210, on the other hand, is even more isolated than Leo A, and
external tidal effects from the MW and M31 cannot have played a
significant role in its evolution (Equation~\ref{tff}). Interactions
are not likely to have affected its observable properties, and so its
unusual SFH must be explained without recourse to interactions. The
total mass of DDO210 is important to consider as well, since its
circular velocity is $\sim 16$\,km\,s$^{-1}$, implying a mass of $\sim
10^{7 - 8}$\,M$_\odot$ (\citealt{begum2004}). DDO210 is an
intrinsically low mass galaxy which has not interacted with large
masses, and yet it did not form most of its stars until $\sim 4$\,Gyr
ago.

What seems clear is that DDO210 must have accreted its gas prior to
reionisation, since it is far too low mass to successfully accrete gas
post-reionisation. Since the virial temperature of DDO210's halo will
not have been vastly higher than the temperature of the accreted gas,
the gas will probably have remained in a relatively diffuse
configuration in the dark matter halo. What star formation occurred at
these early times (if any) will have been small and will have been
confined to localised gaseous overdensities, rather than acting on a
global scale. Deeper observations are required to determine the exact
fraction of old stars in this galaxy. Since the gas never really
collapsed all the way to the center of this system, then a bona-fide
disk will not have had a chance to form. This would explain why we,
along with others, conclude that it is unlikely that the dominant
structure in DDO210 is a disk.

Why are most stars intermediate age? It could be that DDO210 continued
to accrete mass (but not gas) post-reionisation, such that eventually
the gas collapsed sufficiently that star formation proceeded at a
reasonable rate and formed the dominant intermediate age
component. Alternatively, it may be that the gas eventually collapsed
sufficiently without the overall mass of DDO210 increasing, either
through random perturbations or via normal cooling mechanisms.

The above is, of course, speculation, but a scenario such as this
seems plausible given our results and the current cosmological
paradigm. As well as the dominant intermediate age population,
however, the recent SFH of DDO210 is also unusual. The current SFR in
DDO210 is basically zero ($< 0.000003\,M_\odot\,{\rm yr}^{-1}$;
\citealt{vanzee1997,young2003}), even although this galaxy was forming
stars as recently as a few tens of Myrs ago. We quantify this apparent
inconsistency by assuming DDO210 has formed all of its young stars
within the past 500\,Myrs. We compare the number of young stars
observed in the colour -- magnitude range $\left(V - I\right) < 0.5$,
$I < 24.3$\,mags with the corresponding number in an IAC-STAR
simulation, convolved with our error distribution, which has a
constant SFR over the last 500\,Myrs. In order to match the number of
young stars observed, the average SFR over the last 500\,Myrs must be
$\sim 0.0002\,M_\odot\,{\rm yr}^{-1}$ ie. nearly two orders of
magnitude larger than the derived upper limit on the current SFR. It
could be that the recent SFH of DDO210 is best characterised by a very
low, relatively constant SFR, interrupted by periods of inactivity. This
`gasping' star formation scenario has already been implied for several
other Local Group galaxies (eg. \citealt{tosi1991}).

The young stars in DDO210 are not centered in the middle of the
galaxy, but are offset a few arcmins to the east. This location is
coincident with a slight dent in the HI, and perhaps suggests that
some recent event, such as an interaction, could have triggered
off-center star formation in this galaxy (although what could have
interacted with DDO210 is a mystery!). Alternatively, the off-center
star formation may be due to recently captured gas, which has not yet
become fully virialised.  The fact that the young stars are still
clumped on a scale of $\sim 1$\,arcmin ($\sim 0.3$\,kpc) suggests that
they were probably born with a relatively small velocity dispersion
($\sigma \sim 3 - 6$\,km\,s$^{-1}$). We note that
\cite{dohmpalmer2002} showed that the recent star formation in
Sextans~A was not confined to the center of the galaxy, and seemed to
migrate with time. They argued that star formation in one part of a
cloud could cause star formation in another part of the cloud, by
triggering instabilities in the gas. In this case, no interactions
would be required to explain the spatial location of the young stars,
which would reconcile our finding with the splendid isolation of
DDO210.

Finally, we note that the presence of a spatially extended
intermediate age population and a centrally concentrated young
population is compelling evidence for the presence of multiple
structural components in DDO210. The systematic change in ellipticity
of DDO210 as a function of radius supports this claim. These findings
provide yet more evidence that dwarf galaxies are not simple single
component stellar systems (eg. Sculptor, \citealt{tolstoy2004}; Leo~A,
\citealt{vansevicius2004}; NGC6822, \citealt{demers2005};
Andromeda~II, \citealt{mcconnachie2006b}).

\section{Summary}

We have used deep $VI$ imaging from the Subaru Suprime-Cam wide field
camera, and $BR$ imaging from VLT FORS1, to analyse the stellar
content, SFH and structural properties of the low mass, isolated,
Local Group transition-type dwarf galaxy DDO210. We confirm the
distance modulus of this galaxy to be $\left(m - M\right)_o = 25.15
\pm 0.08$\,mags. Main sequence and blue loop stars are consistent with
having formed within the past $60$\,Myrs, whereas the luminosity of
the brightest AGB stars show that this galaxy must have been forming
stars within the past $3 - 6$\,Gyr. This is consistent with the age
estimate derived from the mean $I$-band magnitude of the red clump,
which shows that the majority of stars have an average age of
$4^{+2}_{-1}$\,Gyr. From the shape of the red clump, we estimate that
any contribution from stars with ages $\ge 7$\,Gyr, if present at all,
cannot amount to more than $\sim 20 - 30$\%. We discuss the
implications of this rather unusual SFH within the cosmological
context of dwarf galaxies. In this respect, DDO210 provides a useful
insight into star formation in low mass haloes, since tidal effects
cannot have affected the evolution of this isolated system.

The young stars in DDO210 have a different radial distribution to the
intermediate age population, and are offset to the east of this older
population. When compared with the spatial distribution of HI, we find
that the highest density of young stars is coincident with an apparent
`dent' in the HI contours. In addition, the strong difference in the
radial distribution of the intermediate age stars compared with the
centrally concentrated young population, supports the idea that DDO210
consists of multiple structural components. Clearly, dwarf galaxies
are not simple stellar systems.

\section*{Acknowledgments} 

We would like to thank the referee, Myung Gyoon Lee, for a careful
reading and helpful comments. This work has made use of the IAC-STAR
Synthetic CMD computation code. IAC-STAR is supported and maintained
by the computer division of the Instituto de Astrofísica de
Canarias. We are grateful to Lisa Young for sending us her HI data
which we have used in this paper, and to Myung Gyoon Lee for providing
their photometry of DDO210 to allow for the comparison between
datasets. AM would like to thank Kim Venn and Jorge Penarrubia for
enjoyable discussions during the preparation of this work, James
Bullock for insight into linking cosmology with star formation
histories, and Sara Ellison and Julio Navarro for financial
support. This work is partly supported by a Grant-in-Aid for Science
Research (No.16540223) by the Japanese Ministry of Education, Culture,
Sports, Science and Technology.

\bibliographystyle{apj}
\bibliography{/pere/home/alan/Papers/references}

\end{document}